\documentstyle[12pt,aasms4,psfig]{article}
\def\IT{Iben \& Tutukov~}
\def\P{Paczy\'{n}ski~}

\def\vdh{van~den~Heuvel}
\newcommand{\kms}{\mbox {km s$^{-1}$}}
\def\apgt{\ {\raise-.5ex\hbox{$\buildrel>\over\sim$}}\ }
\def\aplt{\ {\raise-.5ex\hbox{$\buildrel<\over\sim$}}\ }

\newcommand{\pyr}{\mbox {{\rm yr$^{-1}$}}}
\newcommand{\rs}{\mbox {$R_{\odot}$}}

\newcommand{\ms}{\mbox {$M_{\odot}$}}
\newcommand{\ace}{\mbox {$\alpha_{ce}$}}
\newcommand{\md}{\mbox {$\dot{M}$}}

\newcommand{\mc}{\mbox {$M_{Ch}$}}

\newcommand{\at}{\mbox {$\alpha_{th}$~}}

\newcommand{\myr}{\mbox {~${\rm M_{\odot}~yr^{-1}}$}}

\def\m{^m\kern-7pt .\kern+3.5pt}
\def\y{^y\kern-1.9mm .\kern+.3mm}
\def\s{^s\kern-1.7mm .\kern+.3mm}
\def\hup{^{h}\kern-2.1mm .\kern+.6mm}
\def\d{^{d}\kern-2.1mm .\kern+.6mm}

\def\etal{{et al.}\ }

\def\apj{ApJ}

\begin{document}

\title{Type Ia Supernovae: An Examination of Potential Progenitors 
and the Redshift Distribution}

\author{{\it L.~Yungelson}\footnote{Permanent address: Institute of
Astronomy of the 
Russian Academy of Sciences, 48 Pyatnitskaya Str., 109017 Moscow, Russia}
and {\it M.~Livio}\\
~\\
Space Telescope Science Institute\\
~\\
3700 San Martin Drive, Baltimore, MD 21218}

%\maketitle

\begin{abstract}

We examine the possibility that supernovae type~Ia (SN~Ia) are produced
by white dwarfs accreting from Roche-lobe filling {\it evolved}
companions,  under the assumption that a strong optically thick stellar 
wind from accretor is able to stabilize the  mass transfer.
We show that if a mass transfer phase on a
thermal timescale precedes a nuclear burning driven phase, then such 
systems (of
which the supersoft X-ray sources are a subgroup) can account for about
10\% of the inferred SN~Ia rate.

In addition, we examine the cosmic history of the supernova rate, and we
show that the ratio of the rate of SN~Ia to the rate of supernovae
produced by massive stars (supernovae of types II, Ib, Ic) should
increase from about $z = 1$ towards lower redshifts.

\bigskip
\noindent Subject headings: binaries: close-binaries: symbiotic-stars:
mass-loss-stars: novae-supernovae: general-X-rays: stars-galaxies:
evolution

\end{abstract}

\newpage

\section{INTRODUCTION}

Type Ia supernovae (SN Ia) are believed to result from the thermonuclear
disruptions of
accreting C-O white dwarfs (WDs) in binary systems. However, their
immediate progenitors  have not yet been identified. Two main scenarios are
currently discussed most frequently. In one, the exploding white dwarf
reaches the Chandrasekhar mass, $\mc \approx 1.4 \ms$, and carbon
ignition occurs in the WD's center. In the second,
the accreted layer of helium on top of a C-O WD ignites off-center,
at sub-Chandrasekhar masses (edge-lit detonations, henceforth, ELD).
For the most recent reviews of the different aspects of the
evolutionary scenarios which may result in
potentially explosive configurations see e.g.\ Branch et~al.\  (1995),
Wheeler (1996), Livio (1996a), Ruiz-Lapuente,
Canal, \& Burkert (1997),  Yungelson \& Tutukov (1997).
 
The double degenerate scenario (Webbink 1984; Iben \& Tutukov
1984), in which the exploding object is of a Chandrasekhar mass or higher, 
is one of the two evolutionary channels
for which estimates based on population synthesis studies 
(see Yungelson \& Tutukov
1997 and references therein) give occurrence rates of SN Ia of the same order 
as those inferred for the Milky Way Galaxy: $\sim 10^{-3}$ \pyr\
(Cappellaro et~al.\  1997). 
This channel is able to account for the occurrence of SN Ia both in young and
old stellar populations. In the second channel, 
the accretion of helium from a Roche-lobe filling
nondegenerate helium star may result in an
ELD (Livne  \& Glasner 1990; Woosley \& Weaver 1994; Livne
\& Arnett 1995) in sub-Chandrasekhar
WDs at comparable rates. However, this channel may operate only in relatively
young populations (age $\aplt 10^9$ yr), and it is not clear if it can account for the high-velocity intermediate-mass
elements that are seen in SN~Ia spectra (see e.g.\ Branch et~al.\  1995 for a discussion). The double degenerate scenario 
on the other hand is plagued by two basic weaknesses. 
First, no close binary white
dwarf system with a total mass higher than \mc\ has been found yet.  
While Iben, Tutukov, \& Yungelson (1997)  have shown that 
the fraction of close binary 
carbon-oxygen white dwarfs with $M_1+M_2 \apgt \mc$ 
among the potentially observable pairs
hardly exceeds $\sim 1/200$, observations are beginning to place 
meaningful constraints on this scenario (e.g.\ Saffer \& Livio 1997).

Second, existing calculations of the merger process (e.g.\ Benz et~al.\ 
1990; Mochkovich \& Livio 1990; Mochkovich, Guerrero, \& Segretain
1997; Segretain, Chabrier, \& Mochkovich 1997) tend to indicate that the 
outcome of the merger may be an accretion induced collapse rather than an
explosion, or the formation of a single white dwarf.
The observational counterparts of the systems in which the accretion of
helium from nondegenerate stars occurs are also not known, although hot blue
subdwarfs are prime candidates. 

The circumstances listed in the preceding paragraphs stimulated the search 
for other channels which could lead to a sufficiently high occurrence
rate of SN Ia.  In particular, the interest in 
the systems in which white dwarfs accrete hydrogen at high rates, 
as potential progenitors, has 
increased substantially (e.g.\ Di Stefano 1996; Livio 1996a; 
Yungelson \etal\ 1996; Di~Stefano \etal\ 1997; Li \& van den Heuvel 
1997). These systems may show up as supersoft X-ray sources
(see Kahabka \& van den Heuvel 1997 for a review). 
Recently, Hachisu, Kato, \& Nomoto (1996, henceforth HKN) suggested 
what they termed as ``a new model for the progenitor
systems of type Ia supernovae''. This model envisions a massive white dwarf
accreting from a Roche-lobe filling, well evolved ($M_{He~ core} \apgt 
0.35 \ms$), low-mass red giant. In fact,  such configurations were 
suggested as potential SN Ia progenitors as early as 1973 (Whelan \& Iben 1973) 
and were considered by Iben \& Tutukov  (1984) and Yungelson et~al.\  (1996, 
henceforth, YLTTF). The latter authors have shown that
in this model the rate of explosions of Chandrasekhar mass WDs is
more than an order of magnitude below the desired one. However, HKN argued
that with certain modifications to the input physics, this model can produce 
a Galactic rate of SN~Ia of $\sim 0.002$ \pyr\ (although this number was 
obtained as a rather crude estimate,
without a detailed modeling of the underlying population). 

In the present work we incorporate the modifications of HKN 
into a population synthesis code and thereby estimate the
occurrence rate of SN Ia  due to this mechanism. We show that even under the 
most favorable assumptions,  this channel is not
able to produce more than $\sim 10^{-4}$ events per year. We suggest however 
some modifications to the original HKN scenario which somewhat 
increase its efficiency.
Our implementation of the HKN model and the suggested modifications are
described in \S2.  The results are presented and discussed in \S3.  In
\S4 we examine the question of SN~Ia from a cosmic perspective, in view
of recent findings on the history of the star formation rate in the
universe.  Our conclusions follow.

\section{THE MODEL}

\subsection{The Hashisu-Kato-Nomoto model}

The essence of the HKN model may be described as follows. As it was noticed 
by Refsdal \& Weigert (1970, 1971), the radii, luminosities, and core
growth rates of stars with degenerate helium cores depend mainly on the core
mass itself. The same is true for stars with degenerate carbon-oxygen cores
(\P 1970, Uus 1970). This allows to
follow the evolution of thermally stable subgiant  stars, both underfilling and
filling their  tidal lobes, without detailed evolutionary computations. 
The stars (of mass $M_2$ and radius $R_2$) which continuously fill
their Roche-lobes have to satisfy the following criteria:
\begin{equation}
R_{2} = R_{cr}~~~~~ {\rm and}~~~~~
\frac {d \ln R_{2}}{d \ln M_2} = \frac {d\ln R_{cr}}{d \ln M_2},
\end{equation}
where $R_{cr}$ is the radius of the Roche-lobe.
If some of the matter which is transferred from the donor to the 
accretor is lost by
the latter, taking away the specific angular momentum of the accretor,
Eq. (1) gives for the mass loss rate by the donor [this Eq.\ is
equivalent to Eq.~(5) of HKN]: 
\begin{equation}
\frac {\md_2}{M_2} = - \frac {1}{2} \frac {\dot R}{R} \times
\left[1 - \alpha q -\frac {1}{2} \frac {1 - \alpha}{1 + q} q 
- \frac {1 - \alpha}{1 + q} q^2 -
\frac {1}{2} (1 + \alpha q)  f^{\prime} \right]^{-1}.
\end{equation}
Here $q = M_2/M_1$ is the mass ratio, $f^{\prime}$ is the
logarithmic derivative of the relative radius of the Roche lobe,
and $\alpha$ is the fraction of the transferred matter which is
retained by the accretor (retention efficiency).  If $\alpha$ is
close to 1, Eq.~(2) is valid for $q \aplt 0.78$.  For larger
values of $q$, the denominator in Eq.~(2) becomes negative and
the expression gives a positive value of~ $\md_2/M_2$.  This
means that under these conditions mass transfer proceeds on a
thermal or dynamical timescale (e.g.\ Hjellming \& Webbink 1987;
Soberman, Phinney, \& van den Heuvel 1997).  In the conventional
model of evolution, this implies that the accretion rate exceeds
the maximum rate at which the WD can burn the accreted hydrogen
stably, $\dot M_{cr} \sim 10^{-7}$ \myr.  A common envelope
forms, engulfing the entire system, and the subsequent evolution
may lead to the formation of a double degenerate system or to a
merger of the components. For binaries which satisfy the
criterion $q \aplt 0.78$, Eq.~(2) was applied by e.g.\
Yungelson (1973), Webbink, Rappaport, \& Savonije (1983), Iben
\& Tutukov (1984), Kraicheva et~al.\  (1986), de Kool, van den
Heuvel, \& Rappaport (1986), and Yungelson et~al.\  (1995, 1996)
to a number of problems related to the evolution of Algols,
low-mass X-ray binaries, symbiotic stars, and supersoft X-ray
sources.

HKN suggested that an optically thick stellar wind from the
white dwarf (Kato \& Hachisu 1994) may stabilize the mass
transfer for systems with $q \apgt 0.78$.  An optically thick
wind may blow off the excess of matter above $\dot M_{cr}$ at a
rate of up to $\sim 10^{-4}$ \myr. The velocity of the wind may
be as high as $\sim 10^3$ \kms\ and its specific angular
momentum is that of the accretor. If an optically thick wind
indeed operates, $\alpha$ may become close to 0.  In such a
case, Eq.~(2) becomes valid for $q \aplt 1.15.$ This could in
principle increase the number of systems which undergo stable
mass transfer.  From an evolutionary point of view it should be
noted that frictional drag  inside the wind is not able to
significantly affect the orbit of the companion star (e.g.\ Kato
\& Hachisu 1994).

HKN performed a set of test integrations of (their version of) Eq.~(2)  
for a number of combinations of donor and accretor
masses. They have shown that in  systems with initial donor masses of 
$M_d \approx 0.8 - 1.4$ \ms, (with He
core masses $M_{He0} \ge 0.35$ \ms), and accretor masses $M_a \approx
1.0 - 1.2$ \ms,  which obey the criterion $q = M_d/M_a \leq 1.15$,
the accretors can actually grow in mass to \mc.

\subsection{Modifications to the Hachisu-Kato-Nomoto model}

Let us introduce $\xi_L = \frac{d \ln R_2}{d \ln M_2}$~for the mass 
derivative of the
Roche-lobe radius,  $\xi_{ad} = \left( \frac {\partial {\ln R_2}} 
{\partial {\ln M_2}} \right)_{ad}$~
 for the adiabatic derivative of the stellar radius and
$\xi_{th} = \left( \frac {\partial {\ln R_2}} 
{\partial {\ln M_2}} \right)_{th} $ for the derivative of the radius of the 
star in thermal equilibrium.
Dynamical stability of the mass exchange process requires $\xi_L <
\xi_{ad}$, and thermal stability requires $\xi_L < \xi_{th}$  (e.g.\  Ritter
1996). For fully convective stars $\xi_{ad} = -1/3$.
However, Hjellming and 
Webbink (1987) have shown, by an analysis of polytropic models,
that in the presence of a condensed radiative 
core, $\xi_{ad}$ may  be significantly higher than $-$1/3. 
Giant branch stars with masses of degenerate cores 
$M_{He} \apgt 0.25$ can be adequately approximated by polytropes with
condensed radiative cores and convective envelopes. Stars with less
massive cores have relatively shallow convective envelopes and for them
the criteria for stability derived for stars with radiative envelopes are more
appropriate. In Fig. 1 we plot the dependence of $\xi_L$ on $q$
for two extreme values of the retention efficiency $\alpha$ = 0 and 1.
As an example (appropriate for the present study), we compare
$\xi_L$ with  $\xi_{ad}$ for two values of the
fractional mass of the condensed core: $m_c \approx 0.2$ and $m_c \approx
0.39$ (taken from Hjellming \& Webbink 1987). 
It is clear that for values of $\alpha$ close to
0, Roche-lobe filling giants may be dynamically stable 
even for values of $q$ much higher than 1.15. However, if $\xi_{th} < 
\xi_L < \xi_{ad}$ these stars are thermally unstable. 
For giants with radii and luminosities independent of mass, $\xi_{th} = 
0$. This analysis suggests  to include as candidate SN~Ia progenitors also 
stars which lose mass on a thermal timescale. This means that we have to 
reject from the sample of candidate systems only those
systems whose donors had $M_{He} > 0.25$ \ms\ and $\xi_L > \xi_{ad}$
(for the appropriate values of $q$ and $m_c$).

Existing calculations of the mass exchange in low-mass binaries with  donors 
possessing a deep convective envelope do not provide sufficient
information on
the time-dependence of $\md_2$. We therefore applied the following simplified algorithm 
(see also Rappaport, Di~Stefano, \& Smith 1994).
We estimated the mass loss rate,   $\md_{th}$, 
which corresponded to the thermal timescale
of the donor at the instant of Roche Lobe overflow (RLOF) and assumed that the star loses mass at
a constant rate $\md_2 = \alpha_{th} \md_{th}$  
until Eq.~(2) becomes valid. Several sets of runs were performed 
for values in the range $0.1 \leq \alpha_{th} \leq 2$. Systems with initial 
transfer rates of $\vert \md_2 \vert > 10^{-4}$ \myr\
were rejected, because for them, optically thick wind solutions do not exist (Kato \&
Hachisu 1994).   
   
In addition, we calculated the occurrence rates of events in which
ONe white dwarfs reached $M_{Ch}.$~ While these events are expected to
produce accretion induced collapses (AIC) rather than SN~Ia 
(Canal \& Shatzman 1976), they are of great interest for the
formation of low-mass X-ray binaries. In our population synthesis
code, the ONe white dwarfs are assumed to descend from stars with MS masses
between 9 and 11.4 \ms. The width of this mass interval may be somewhat
overestimated, however it is quite clear that stars with masses of 
10--10.5~\ms\ do produce ONe
white dwarfs (e.g.\ Dominguez, Tornamb\`e, \& Isern 1993;
Ritossa, Garcia-Berro, \& Iben 1996). 
The minimum mass of an ONe white dwarf in our code is 1.19 \ms.  
Actually, cold and massive  ($M \apgt 1\,\ms)$\
CO white dwarfs may also experience 
AICs instead of supernova explosions (e.g.\ 
Nomoto \& Kondo 1991;  Canal \& Gutierrez 1997). The mass range
of CO white dwarfs able to collapse is strongly dependent on the
approximations adopted for, e.g.\ the thermal structure of the WD and the
nuclear reaction rates. However, with this in mind, we have to note that
the rate of events considered below as SNe~Ia may be reduced in favor
of AICs.
 
The initial set of systems to which the mass loss algorithm was applied was
generated by a population synthesis code which has already been used
for a number of problems concerning binary star evolution. The detailed 
description of all the assumptions involved in the synthesis as well as 
approximations for the dependence of stellar
parameters on mass can be found, for example, in Yungelson et~al.\  (1996)
and references therein. The occurrence rates of events were obtained by 
means of  Eq.~(1) for the binary birthrate from \IT (1984). The same 
equation was used for the estimates of HKN and Li \& van den Heuvel 
(1997).  

\section{RESULTS AND DISCUSSION}

\subsection{SN~Ia in systems with low-mass giant donors}

As a first step in our modeling, we generated a population of systems which
satisfy the criteria for dynamically stable mass exchange.
Every system in this sample was then followed throughout its
evolution. The values of $\xi_{ad}$\ as a function of the relative mass of the
core were obtained from Hjellming \& Webbink (1987) and the retention
efficiency $\alpha$\ was estimated at each time step as $\md_{cr} /
\md.$ For $\md_{cr}$\ we adopted the data from Iben \& Tutukov (1989).
    
If initially the donor stars were thermally unstable,
then they were stripped of mass until Eq.~(2) became valid, according to
the prescription given in \S2.2. For stars  which were able to evolve on 
a nuclear timescale initially, or entered this phase after a  ``thermal'' 
timescale mass loss phase, we used the dependence of 
the radius, luminosity, and core growth rate on the mass 
of the core (in solar units and years), derived by
Iben \& Tutukov (1984) for solar chemical composition stars:\footnote[1]
{Here we correct a misprint in the power of the $L(M_{He})$ dependence 
in Yungelson et~al.\  (1996)}
\begin{equation}
R \approx 10^{3.5}M^4_{He},~~~~~~
L \approx 10^{5.6} M^{6.5}_{He},~~~~~~
\dot M_{He} \approx 10^{-5.36} M^{6.6}_{He}. 
\end{equation}
Combining the first and the third expressions in (3) with Eq.~(2),
we obtained the mass loss rate and the timescale of the evolution 
(which is controlled by nuclear burning).  Formulae~(3) give values 
of \md\ which are in very good agreement with 
those obtained by Webbink, Rappaport \& Savonije (1983; the latter were used by HKN). The integrations were continued up to 
the point when one of the following situations was encountered. 

(i) The hydrogen-rich envelope of the dwarf was exhausted, it became a He white
dwarf, but the accretor did not succeed to reach  \mc\ or to
accumulate  a ``critical"  mass of He [see option (iii)].

(ii) The accretor reached \mc. Depending on the initial mass of the accretor
this event was classified as a SN~Ia or an AIC.

(iii) A ``critical" mass of He, equal to $\Delta M_{He} = 0.15$\,\ms, was
accumulated on top of the accretor due to hydrogen burning, 
and the accretion rate at
that instant was below $3 \times 10^{-8}$\,\myr. Such events were 
classified as ELDs, which may also be responsible for SN~Ia (e.g.\
Livne \& Glasner 1990; Woosley \& Weaver 1994; Livne \& Arnett 1995).
If the accumulation of 0.15\,\ms\ of He occurred at a higher \md, we 
assumed that the ensuing flash was not accompanied by a significant
mass loss (Kato, Saio, \& Hachisu 1989). Note, that this assumption is 
at variance with that made in our earlier papers, where we prescribed some 
mass loss to accompany such flashes (see e.g.\ Fig.~2 in Yungelson et~al.\  
1995). The reason for the present assumption was to
make our model as similar as possible to that of HKN.

Table 1 summarizes the results of several runs performed for different sets of
assumptions. Note that in the Table, we have labeled the events of accumulation 
of $0.15 M_{\odot}$
as ELD even though the actual outcome is not certain.
Models~1 to 5 correspond to the case in which we followed the HKN
prescription for the selection of candidate systems: only white dwarf plus red
giant systems with a positive
denominator in Eq.~(2) for $\alpha = 0$ were considered for subsequent evolution. 
Model~1, is in fact a strict reproduction of the 
HKN model. No ``explosive" events (of
any kind) were obtained in this model. The reason can be understood by the
following estimate. Systems,
which have at  
RLOF $q \aplt 1.15$ must contain relatively massive white dwarfs.
A $\sim 1$ \ms\ white dwarf descends from a main sequence star of 7--8~\ms.
This means that in the common envelope stage (in which white dwarf was formed),
the separation of the components had
to decrease by a factor $a_f/a_0 \sim  \ace (M_{WD}/M_1)(M_2/M_{env}) \sim
1/40 - 1/50$,
if $M_2 \sim 1$ \ms\ and \ace=1 [we apply for this estimate
Eq.~(17) of Iben and Livio (1993)
for the variation of separations in common
envelopes; here $a_f$ and $a_o$ are the final and initial separations, 
respectively, and
\ace\ is the common envelope efficiency parameter].
The widest, but still close in the evolutionary sense,
systems with 7~\ms\ primaries and 1  \ms\ secondaries have orbital
separations of $\sim 1500$ \rs. After the first common envelope stage
this separation is reduced to $a_f \sim 38$ \rs.
Subsequently, at RLOF, the radius of
the donor is $\sim 14.5$ \rs\ and $M_{He} \approx 0.26$ \ms.
However, Roche-lobe filling stars with such
small helium cores lose mass at rates which hardly ever exceed 
$\sim 10^{-7}$\myr\ (see e.g.\ Fig.~1 in Kraicheva et~al.\ 1986). 
Consequently, instead of accumulating matter (by steady hydrogen burning),
the accretors enter a regime of thermonuclear runaways and are unable to
grow in mass. In the model calculations, the most
massive cores of the donors at RLOF did not exceed $\sim 0.3$ \ms. 
{\sl This means
that in its original form, the HKN model (if combined with a value of 
\ace=1, which we consider as ``standard",
e.g.\ Rasio \& Livio 1996) is unable to produce  any SN~Ia.} 

Our choice of \ace=1 as a ``standard" value is based on the success 
of attempts to model, by
means of a population synthesis, the numbers and main parameters
of such different constituents of the binary
star population of the Galaxy as Wolf-Rayet stars, massive X-ray binaries,
symbiotic stars, cataclysmic variables, double degenerates of different kinds
and to obtain reasonable agreement with observations.                   
However, the estimate of $a_f/a_0$, which is based on rough energy budget considerations 
may not be very accurate. Furthermore, different formulations 
of the equation for $a_f/a_0$ are
available in the literature and as a result, even the value of
\ace\ = 1 may correspond to different amounts of energy deposited into
the common envelope in different studies (see Livio 1996b for a discussion). 
Therefore, it is worthwhile to 
study the influence of  variations in \ace\ on the rate of SN~Ia.

It is evident that a reduction in \ace\ would result
in an even more severe upper limit on the masses of He cores
at RLOF. Results of varying \ace\ between 1 
and 20 are given in rows~2
to 5 of Table~1. Clearly, the conditions for accumulation of \mc\ become more
favorable with the increasing range of the masses of He cores of the
donors. Even for \ace=2 some SN~Ia appear. However,
even in the relatively ``generous" case of \ace=10
the occurrence rate of SN~Ia is
still two orders of magnitude below the inferred Galactic rate. The upper 
panel of Fig.~2  shows the 
position of successful systems in the plane of the initial 
mass of the accretor $M_a$ - initial mass of the donor $M_d$, for 
$\alpha_{CE} = 10$, in the HKN model. The figure demonstrates 
that actually, the original HKN model, if it works at all, is favorable
not for SN~Ia but for AICs.

Rows~6 to 9 of Table~1 show results of computations 
for the model of HKN modified
to take into account the initial ``thermal" stage of mass exchange, as described
in \S2.2, for a number of scaling factors for \md. The immediate
result is the appearance of SN~Ia even for \ace = 1, for which the original HKN
model failed. The results for our version of the model are clarified by 
Fig.~3, in which we plotted the evolution of the donor and accretor masses for 
a 1.75 + 0.85 \ms\ system, 
with $a_0 = 10$ and 20 \rs, for $\alpha_{th} = 0.5$.   In the first case, 
the mass of the He core
of the donor at RLOF is $\sim 0.19$ \ms. Therefore, under our assumptions, 
this system is not subject to the restrictions imposed by the requirement of dynamical stability
and the donor begins to experience mass loss on a thermal timescale. 
For $\alpha_{th}$ = 0.5 the mass transfer rate is $\sim 6.6 \times 10^{-7}$ 
\myr. For this value of \md\ and with $M_a = 0.85$~\ms\ the initial
retention efficiency is $\alpha \sim 0.6$, i.e.\ 40\% of the accreted matter is
blown off from the system. With increasing $M_a$, the retention efficiency also
increases and it becomes equal to unity when $M_a \approx 1.225$ \ms. The total mass lost from
the system by the optically thick wind is only about 0.1~\ms. The mass of the
donor grows to \mc\ very shortly before Eq.~(2) becomes valid, still in the
``thermal" stage of mass exchange.

If $a_0 = 20$ \rs, the radius and luminosity of the donor at RLOF are higher
and for $\alpha_{th}$ = 0.5 the initial mass transfer rate is $\md \approx 2 
\times 10^{-6}$\,\myr. The initial retention
efficiency is only $\sim 0.2.$ The mass of the donor drops faster than in the
former case, but the rate of growth of $M_a$  is the same, because it is limited by
the maximum stable hydrogen burning rate. When 
the component masses reach the values of
$M_d \approx 1.05$\,\ms\ and $M_a
\approx 1.0$\,\ms, Eq.~(2) becomes valid (formally discontinuously). \md\ is
initially very slightly above $3 \times 10^{-8}$ \myr. At this \md, the
accretor is still able to grow in mass despite nova explosions
(according to the calculations of Prialnik \& Kovetz (1995)
the results of which are incorporated into our code). The retention
efficiency is close to 0.05. At the instant
when \md\ drops below $3 \times 10^{-8}$\,\myr, the mass of the accretor is
1.004~\ms\
and, formally, the criterion for an ELD is satisfied. It is important to note that
most of the mass of the ``critical" He layer is accumulated at high \md.

The lower panel of Fig.~2 shows the position of systems which  accumulate \mc\ 
in the initial $M_a - M_d$ plane,
in a modified HKN model, with $\alpha_{th} = 0.5,~ \ace = 1$.  
It is clear from this Figure that the 
systems which are able to produce SN~Ia are those which experience a thermal
timescale mass exchange, i.e.\ systems, which are slightly evolved off the 
main sequence. This suggests again that supersoft x-ray sources, 
in which relatively massive white dwarfs accrete from evolved companions (van 
den Heuvel et~al.\  1992; Rappaport, Di~Stefano \& Smith 1994; YLTTF; Kahabka 
1995; Southwell et~al.\ 1996; Li \& \vdh\ 1997; Kahabka \& \vdh\ 1997; Di~Stefano 
1996; Di~Stefano et~al.\ 1997) should be closely examined.

The main difference between the present model and the models 
considered by Rappaport et~al.\ (1994) and YLTTF is in 
the fact that we allow for mass transfer rates above the limit for stationary
hydrogen burning by assuming that the excess mass leaves
the system and stabilizes the mass transfer. This results in
wider ranges of initial mass ratios of the components and 
masses of the cores of the donors. The net effect of this difference is that 
in about 15 times more systems the white dwarfs can  reach \mc,
compared to previous calculations
(compare e.g.\ models~6 and 10, the latter of which was computed
under assumptions similar to those 
in YLTTF). It is important to note in addition that all the white
dwarfs which grew to \mc\ in the YLTTF
model were of the ONe variety, while in the present model CO
white dwarfs also grow to \mc.

A comparison between models 10 and 11 shows the effects of varying 
\ace, under 
assumptions that are similar to those in the YLTTF model.  

A comparison of models 6~and 9 shows the effects of varying the
scaling  factor for the mass loss rate during the
``thermal" stage. There exists a value of $\alpha_{th}$  for which the
number of ``explosive" events reaches a maximum. If $\alpha_{th}$ is high,
i.e.\ \md\ is high, the retention efficiency is close to zero, but after 
the mass ratio drops below $\sim 1.15$, 
the accretion rate also decreases sharply and 
either a situation which is favorable for an ELD is encountered or an
erosion of the already accumulated He layer begins. If $\alpha_{th}$
is low, the white dwarf may be eroded 
from the very beginning of the accretion process.
A comparison with Fig.~1 of Li \& \vdh\ (1997), which shows an example
for the behavior of \md\ {\em 
vs.}\ $M_d$~ from their evolutionary calculations, suggests that \md\ 
may be initially higher than obtained in our simplified models. However, as 
our trial computations for \at = 2 (not included in Table~1) have shown,  
more mass is then lost in the stage with virtually zero retention 
efficiency and the rate of explosive events decreases as compared to the 
\at = 1 case.

The criteria for dynamically and thermally stable mass transfer discussed in
\S2.2 were derived under the assumption that the stars may be represented by
polytropic models. While in general this assumption proves to be quite
reasonable, it is worthwhile to test its influence on the model. 
Model~12 gives the results of a trial run with a very ``restrictive"
model, in which we required that the mass exchange has to be dynamically stable
irrespective of the fractional mass of the He core, i.e.\ the restriction
$\xi_L < \xi_{ad}$ was imposed on all the candidate systems. This restriction
efficiently eliminates all the systems with accretors able to grow to \mc.

In contrast, in models 13 to 16 we did not impose any
restrictions on the candidate systems. Clearly, these models are the most
prolific producers of explosive events. For $\alpha_{th}$ = 0.2 a
SN~Ia rate which is consistent with observations
is obtained.  Varying $\alpha_{th}$ has the same effects on
these models as those discussed above for models 6--9.  However,
limitations based on the stability of the models have to exist. 
Trial computations with a full-scale Henyey code (A.~Fedorova, private 
communication) for a 2.0 + 1.2 \ms\ system, with $M_{He0} = 0.4$ \ms\ and 
a zero retention efficiency show that the mass transfer rate increases to 
$\sim 3 \times 10^{-3}$ \myr\ in 80,000 yr after RLOF. The same behavior is 
suggested by the computations of Li \& \vdh\ (1997).
This behavior is in complete agreement with the results of
Hjellming \& Webbink (1987): $\xi_{ad}$ becomes negative when the 
fractional mass of  the core exceeds $\sim 0.21$. 
In fact, in our model we rejected such systems because the
initial ``thermal" \md\ was in excess of $10^{-4}$ \myr.
The systems  with component masses of 2 and 1.2 \ms\ 
in Fig.~2{\it b} have lower fractional core masses.

Runs 13--16 provide  the upper limits for the possible rates of
SNe~Ia and ELDs from the model under consideration, since the only formal
restriction applied was that the mass loss rate by the donor should be
$\md \leq 10^{-4}\,\myr,$\ well over the 
Eddington critical accretion rate for WDs. These limits are
valid irrespective of the mechanism which actually removes the excess of
``unaccretable'' matter from the system.
  
\subsection{Comparison to the work of Li \& \vdh\ and Di~Stefano
et~al.}
Li \& \vdh\ (1997) made a series of evolutionary  computations using a 
simplified Eggleton (1971) code, for accretor masses of 1 and 1.2~\ms\ and 
a range of donor masses and initial orbital periods of the system. They 
suggest that the white dwarfs may grow in mass to \mc\ if $0.9 \aplt M_a/\ms 
\aplt 1.2$, $2 \aplt M_d/\ms \aplt 3.5$, $0.5 \aplt P_{orb}/ $day$ \aplt 6$ or 
$M_a \approx  1 \ms$, $0.9 \aplt M_d/\ms \aplt 1.2$, $70 \aplt P_{orb}/ $day$ \aplt 
700$. The shorter period range found by Li \& \vdh\ overlaps with the 
range found by us, although it is slightly shifted to higher 
masses of both the accretors and donors (we do not consider here  
donors with $M_d \ge 2.5$\,\ms, since these stars do not have degenerate 
helium cores and Eq.~(2) cannot be applied). Li \& van den Heuvel obtained
a rough estimate for the rate of SNe~Ia of $2 \times 10^{-3}$\,\pyr. However, 
they did not consider the evolution which leads to the formation 
of high-mass white dwarfs with low or moderate mass companions. If this is 
taken into account, the {\sl total\/} birthrate of massive white dwarf~+ 
main sequence systems, irrespective of their fate,  in the above
range of $M_a$, $M_d$, and $P_{orb}$ and for \ace =1 is  only $2 \times 
10^{-4}$\,\pyr. 
This gives an actual {\sl absolute\/} upper limit to the rate of explosive 
events from the systems considered by Li \& \vdh. Considering their 
long-period range, systems in that range do not form at all because of the 
sharp decrease in the orbital separation in the common envelope stage (i.e.\   
for the same reason which makes the original HKN model ineffective).

The upper limit of the orbital period range of  successful progenitors of 
SN~Ia found by Li \& \vdh\  is slightly below the limit found in the present 
paper (6 to 11~days, depending on the masses of the donor and accretor). This 
may mean that the stabilizing effect of the large fractional cores of the 
donors  may be weaker than Hjellming \& Webbink (1987) suggest. In such a case, 
our models actually overestimate the potential 
occurrence of SN~Ia (see the discussion of model~12 above).

Di~Stefano et~al.\ (1997) addressed the question of SNe~Ia from
low-mass binaries by performing Monte-Carlo simulations of 
the underlying population and evolving it using parameterized results of
Henyey-type computations for the estimate of $\xi_{ad}$\ as a
function of the relative core mass of the donor.  The
retention efficiency was treated by them as an  ajustable parameter
and mass loss rates higher than $\md_{cr}$\ were allowed. The
mechanism for the removal of excess matter was not specified. The
results of Di~Stefano et~al.\ are in complete  agreement with ours: the
rate of SNe~Ia is between $9 \times 10^{-5}$\ and $1.8 \times
10^{-4}$~\pyr\ (for 100\% initial binarity), only a few of percent of
the desired rate. The most favored combination of accretor {\em vs.}\ donor 
for the production of SNe~Ia is $M_a \sim 0.9 - 1.0~\ms, M_d \sim 2~\ms$\ 
(compare to Fig.~2).  In the most ``prolific'' case for the production of 
SNe~Ia, when it is assumed that the binaries under consideration may survive 
the common envelope stage, the supernovae rate is $\sim 0.55 \times 10^{-2}~ 
\pyr$, somewhat higher than in our most prolific case~15. The reason for this 
slight difference may be the inclusion of donors with $\md \apgt 10^{-4}~\myr.$

The rate of events involving the accumulation of 0.2~\ms\ of He on top of the 
WD in the study of Di~Stefano et~al.\ is in all cases higher (by about an
order of magnitude) than the rate of events reaching \mc, and it may amount
to a few tens of percents of the ``observed'' SN~Ia rate (we have higher
rates because we consider the accumulation of lower mass of He).  

\subsection{Uncertainties}
We would like now to discuss a few important uncertainties. 
In the discussion above we considered events in which at least 
0.15~\ms\ of He was
accumulated on the surface of the accreting white dwarf 
and \md\ became lower than $3 \times 10^{-8}$ \myr, as edge-lit
detonations. In fact, this identification represents
somewhat of an extrapolation of our current knowledge of the
nature of these events. In all the existing calculations of He
detonations on top of CO dwarfs, either accretion at a constant
rate was assumed, or a certain mass of He was exploded by a
``piston".  These calculations have shown that for the He
detonation to cause a C detonation, \md\ must be confined to a
rather narrow range of $(1 - 5) \times 10^{-8}$ \myr\ (e.g.\
Woosley \& Weaver 1994).  In the case considered in the present
study,  the rate at which mass is accumulated is that of steady
hydrogen burning. Within the uncertainties which exist in the
estimates of this rate (for white dwarfs with masses below $\sim
1$ \ms), the critical rate for helium detonation may be considered 
as being marginally compatible with the lower limit 
of the stationary hydrogen burning rate $\md_{st}$ (although on the lower side).
However, the mass is actually accreted at a rate which is much higher than  $\md_{st}$.
This may make the effects of compressional heating important for the thermal
history of both the hydrogen and the helium layers. It is thus not entirely clear 
if all the events identified here as ELDs will indeed result in detonations. It should 
also be remembered that there does not exist a unique critical mass for the He layer to 
produce an explosion, rather, it depends (with uncertainties by a factor of a few) on 
the mass of the white dwarf, its temperature, and \md\ (e.g.\ Kawai, Saio, \& Nomoto 
1989;  Iben \& Tutukov 1989; Limongi \& Tornamb\`e 1991; Woosley \& Weaver 1994). 
Another point that should be made is that a formal consideration of Eq.~(2) suggests a 
quite abrupt decrease in \md\ from the high ``thermal" value to the low 
``nuclear" one (the denominator  changes its sign and it is close to 0).
In more realistic models, this transition has to be smooth (see e.g.\ Li 
\& \vdh\ 1997).  However, an important conclusion of the present study is, that 
if the steady accumulation of He is possible and {\sl if it really results in
edge-lit detonations which produce SN Ia}, then semidetached systems with
low mass (sub)giant donors may be the main channel for SN~Ia.  

In the context of the possible contributions of \mc\ accumulations and
ELDs to the rate of SN~Ia it is instructive to consider the age
distribution of the two types of events (Fig.~4). The events, termed by us 
as SN~Ia, happen in systems with massive white dwarfs and relatively massive
companions, and \mc\ is accumulated in them shortly after the donor
leaves main-sequence. Hence, the typical age of these systems has to be
the main-sequence lifetime of a $\sim 1.8 - 2.2~\ms$\ star---of the
order of $\sim 10^9$\,yr. Edge-lit detonations on the other hand, as 
discussed above, occur if the white dwarf accumulates some mass at high \md\ 
(but not reaching \mc) and then \md\ declines to below $3 \times 10^{-8} \myr.$
This requires lower mass donors (typically $\sim 0.6 -0.8~\ms)$, 
lower mass accretors ($\sim (1-2)~\ms)$ and a long stage of nuclear-burning 
powered mass exchange. Hence, these systems have typically to be older. In 
stellar populations with a constant star formation rate the incidence of \mc\ 
accumulations ``saturates'' in $\sim 10^9$\,yr. Their typically younger age 
also suggests that these systems are not the progenitors of SNe~Ia in 
elliptical galaxies, which have short initial bursts of star formation. The 
same is true for the systems producing ELDs, although
their typical ages are in the range of 1 to 6~Gyr. 

In Table~1 we also give the total number of objects under consideration, $N_{obj}$, 
in the Galaxy. All the objects are semidetached systems in which white dwarfs are
either burning hydrogen steadily or are experiencing hydrogen shell flashes of the 
nova type [at the low end of the \md\ predicted by Eq.~(2)]. If the position of the
photosphere in the wind is at a radius smaller than $\sim 0.07$ \rs, the accreting
white dwarf may radiate in supersoft X-rays. The upper limit on the number of
supersoft X-ray sources represented by the entries in col.~(7) of Table~1 is not
significantly different from estimates of the number of Galactic supersoft sources, 
especially if one takes into account absorption effects (e.g.\ Di~Stefano \& 
Rappaport 1994; YLTTF).   

Finally, in col.~(8) of Table~1 we list the number of hydrogen flashes  
on accreting white dwarfs (per year) which may be
expected from the systems under consideration. Note, that we included only
systems which entered the regime of non-steady burning after passing through
a stage of steady accumulation of He. Systems which had $\vert \md \vert 
\leq \vert \md_{st} \vert$ {\it ab~ovo\/} were rejected from the sample because 
they were not interesting as SN~Ia progenitors. The number of flashes was estimated, 
like in Yungelson et~al.\  (1995, 1996), on the basis of models for the accretion 
of hydrogen onto {\it bare\/} CO white dwarfs (Prialnik \& Kovetz 1995). However, the 
existence of the He layer does not change the expected rates significantly (compare 
the critical ignition masses $\Delta m_{acc}$ from  Prialnik \& Kovetz (1995) with 
$\Delta m_{acc}$ for white dwarfs with He surface layers from Prialnik \& Livio (1995); 
the latter are higher only by 30 to 50\% than the former). If we identify 
these flashes with novae, an important property of these novae is that their ejecta 
should be enriched in He, but should have solar metal abundances; such novae are 
known to exist (Livio \& Truran 1994).  It should be noted that some of the 
models in Table~1 produce higher rates of novae than the observed one [even though 
the latter is highly uncertain (e.g.\ Shafter 1997)].  Since the calculated rates 
are definitely lower limits (due to the rejection procedure mentioned above), the 
reality of the associated models should be regarded as questionable.

The uncertainty in the results due to potential differences in the distribution 
of binaries over mass ratios or initial separations is not expected to exceed 
$\sim 20-30$\%, unless an initial mass-ratio distribution which favors low-mass 
companions is chosen (see e.g.\ the study of supersoft X-ray sources by 
Di~Stefano et~al.\ (1994)). As it was shown above, the dependence of the results
on the common envelope parameter may be much more important. Since the most 
prolific producers of SNe~Ia are stars with $M \sim 2~\ms,$ the effects of such
processes as angular momentum loss via magnetic stellar winds are not important 
for them.

\subsection{Supernovae in a cosmic perspective}

Recent progress in the understanding of the cosmic history of the
star formation rate (SFR; Connolly et~al.\ 1997; Madau et~al.\  1996; 
Madau 1997; Pei \& Fall 1995) makes it worthwhile to attempt to gain at 
least a qualitative insight into the evolution of the supernova
rate with redshift $z$. In view of the uncertainties that are still involved,
we approximate the Madau et~al.\ plot for the SFR by two straight lines, which 
gives a peak in the SFR at $z \sim 1.14$. This is almost at the exact position 
of the peak in the SFR as obtained using photometric redshifts (Connolly et~al.\
1997). We should note that the models suggested so far for the cosmic metal 
enrichment rate (or star formation rate) still do not fit well all the points 
in the SFR(z) diagram (see e.g.\ Fig.~3 in Connolly et~al.) 

In our present, highly simplified approach, we have made two crude assumptions.
First, we assume that all the galaxies follow the same SFR history. Second, we 
assume that stellar evolution (which has really been calculated for stars of 
solar composition) is the same in all the galaxies, irrespective of the epoch 
in the history.\footnote[2]{A similar approach was used e.g.\  by Canal, 
Ruiz-Lapuente, \& Burkert (1996) and by Yungelson \& Tutukov (1997).}

Three types of events were considered. First, supernovae
descending from massive
stars: with a main sequence mass $M \ge 11.4$~\ms\ in close binaries 
and $M \ge 10$~\ms\ otherwise. These may be identified with SN II, Ib, Ic 
depending on the presence or absence of a (tenuous) hydrogen envelope at 
the moment of core collapse. The above assumptions on the range of 
progenitor masses give for the present ($z=0$) total
rate of SN II, Ib, Ic a value close to $0.02$ \pyr, which is consistent 
with observational estimates (e.g.\ van den Bergh 1991; Cappellaro et~al.\ 
1997). The second class included mergers of double degenerates, which we still
consider as one of the best progenitor candidate classes for most SN~Ia (see 
Branch et~al.\ 1995 for a discussion). The model produces a rate of mergers 
with a total mass exceeding $M_{Ch}$ in the Galaxy (if a constant SFR for a 
Hubble time is assumed) of 0.003 \pyr, again in agreement with the rates 
suggested for Sbc-Sc type galaxies (e.g.\ Cappellaro et~al.\ 1997). The third 
class included the events of accumulation of 0.15 \ms\ of He on top of white 
dwarfs accreting from nondegenerate helium companions. The rate of such events 
in the Galaxy (under the assumption of a constant SFR) is $\sim 4.5 \times 
10^{-3}$ \pyr, i.e.\ if this channel indeed produces SN~Ia, then it can
account for the majority of the events. The assumption of a constant SFR may 
be reasonable, since the systems in the second and third classes are mostly  
younger than several $10^9$~yr (Yungelson \& Tutukov 1997).  The values of 
$h = 0.5$ and $q_o = 0.5$ were assumed for the cosmological parameters.
  
Fig.~5 gives the history of the above types of SNe as a function of $z$ and 
look-back time. As could be expected, the curve for SN II, Ib, Ic closely 
follows the SFR because the age of these systems is $\aplt 2 \times 10^7$~yr. 
The curve for mergers of double degenerates is somewhat delayed (shifted to a 
lower $z$), because the age of the youngest SN Ia is $\sim 10^8$ yr (i.e.\ the 
age of the typical progenitor stars, of 5--10~\ms, plus the delay before
gravitational waves radiation brings the components into contact).
The systems with nondegenerate helium rich donors in which a potentially 
explosive accumulation of He occurs are again young. The range of masses of 
the primaries in the progenitors of the latter systems partially overlaps
with that for the double degenerates, but the components are typically closer 
after the white dwarf plus nondegenerate  star form. 

While at $z$ close to 1 the helium ignitors dominate (in terms of the rate), at 
$z$ close to 0 the central carbon ignitors (double degenerates) begin to dominate, 
because of the existence of a reservoir of pairs formed at earlier epochs. For the 
systems with nondegenerate helium donors which are candidates for ELD such a 
reservoir does not exist, because it is limited by the lifetime of low-mass  
He-stars, $T_{He} \aplt 10^9$ yr.  

The actual slope of the curves for the incidence of supernovae at $z \sim 1-2$ 
may be slightly less steep than shown in Fig.~5, since our simple approximation for 
the SFR in this interval may be steeper than the actual one.

We also show in Fig.~5 the ratio of the rate of mergers (SN~Ia) to the
rate of explosions of massive stars (SN~II, Ib, Ic).  As can be seen
from the figure, this ratio is nearly constant for $z \apgt 1.2$, but it
increases quite significantly between $z = 1$ and the present.  The
reason for this increase is the build-up of a reservoir of degenerate
pairs in the earlier epochs (higher $z$).

If SNe~Ia were  central ignitors that descend from low-mass semidetached 
systems of the type analyzed in the present paper, then their formation 
history would follow the SFR history (as a function of $z$), because the 
age of these systems is typically below 1~Gyr and no reservoir exists for 
them (Fig.~4; see also Yungelson \& Tutukov 1997). If on the other hand, 
ELDs in low-mass systems really produce supernovae, then their dependence on
redshift or look-back time may be more like that of central carbon ignitors
(due to the merger of double degenerates), because of the existence of a 
considerable fraction of the systems in which ELDs may occur several Gyr 
after formation (Fig.~4).

\section{CONCLUSIONS}

On the basis of the calculations presented in this paper we can draw the
following (somewhat tentative, in view of the uncertainties) conclusions:

1. The incorporation of the possibility of mass loss by the accreting white dwarfs, via
optically thick winds, allows to increase the number of systems which are able
to burn hydrogen steadily. However, if only the nuclear burning-controlled phase of
the evolution of the donors is considered [with $q \aplt 1.15$;
 governed by Eq.~(2)], as suggested by Hachisu, Kato, \& Nomoto
(1996), the rate at which white dwarfs grow in mass to \mc\ does not
exceed a few percent of the currently estimated occurrence rate of Galactic SN~Ia
(see rows 1 to 5 of Table~1).

2. The inclusion of a phase of mass exchange on a thermal
timescale of the donor, which precedes the ``nuclear" phase while $q > 1.15$,
results in an occurrence rate of SN~Ia  of the order of $\sim 10^{-4}$ (see rows 6 to 9 of
Table 1). If the Galactic SN~Ia rate is indeed $\sim 3 \times 10^{-
3}$~yr$^{-1}$, then it appears that this channel
is not the main route to SN~Ia, but it may account for about 10\% of
the events (significant uncertainties still exist however). This may be compatible with the growing suspicion that there could be
several channels of evolution leading to SN~Ia and the possibility that there
exists a diversity in the parameters of the observed supernovae (see e.g.\ Branch et~al.\  1995 for a discussion). 
As potential evidence for the latter suggestion one may consider the differences in
the luminosities of SNe~Ia in volume limited samples (Schaefer 1996).    

3. An interesting property of the model which involves an optically thick wind 
phase is the potential presence of non-negligible amounts of
circumbinary hydrogen. Deep searches for early radio emission
would thus be able to test for the presence of such hydrogen in 
future SNe~Ia (Boffi \& Branch 1995).  

4. One may consider two additional situations in which the accretion
of hydrogen at relatively high \md\ can occur onto white dwarfs, namely, 
in cataclysmic variables and in symbiotic stars. In the former systems 
this may happen just after RLOF by the donor, while in the latter, when 
their donors (losing mass by a wind) are very close to the end of the 
symbiotic star stage (see e.g.\ Yungelson et~al.\  1995, 1996). However, it 
has been shown that the number of events in cataclysmic variables is at the 
level of $10^{-5}$ \pyr\ and it is almost insensitive to variations in the 
criteria for dynamically stable mass exchange and steady hydrogen burning.
The reason for this insensitivity is the fact that for low mass stars, $\md_{th}$ 
is of the order of $10^{-7}$ \myr, and hence, not many systems are rejected from 
the sample because of accretion rates exceeding the maximum allowing for steady 
hydrogen burning.  

Concerning symbiotic stars, Yungelson et~al.\ (1995, 1996) did not consider the 
stage of RLOF, because in an overwhelming majority of the systems, at the beginning 
of RLOF the mass ratios of the components do not satisfy the criteria for dynamically 
stable mass exchange.  However, when these criteria are satisfied,
the mass exchange rate may be as high as $\md\ \sim 10^{-6}- 10^{-5}$ \myr\ (see e.g.\  
de~Kool et~al.\  1986; Pastetter \& Ritter 1989). Consequently, we updated the code to 
take into account the possibility of stable mass exchange and mass loss from the white 
dwarfs via optically thick winds (as described in the present paper). We found, however, 
that the rate of accumulation of \mc\ still remained at the level of $\sim 10^{-6}$ \pyr\ 
and the rate of accumulation of 0.15 \ms\ of He at the level of $\sim 10^{-4}$ \pyr, like 
in the above quoted studies. The reasons for these low rates are: the paucity of systems 
which satisfy the criteria for dynamically stable mass exchange, the short duration of this 
stage (if it happens at all), and the virtual absence of white dwarfs with masses
in excess of 1 \ms\ in the zero-age population of symbiotic systems.

The possible presence of optically thick winds from the white dwarfs in 
symbiotic stars does not influence the number of  systems which go
through two common envelope stages and form double degenerates with
$M_{tot} \geq \mc$ (which later merge due to gravitational wave radiation). 
The reason is that the progenitors of these systems are relatively massive: 
4--10~\ms. As trial runs of the code show, at the second RLOF, despite the
stellar wind mass loss, the companions to the first-born white dwarfs are still
sufficiently massive to experience a dynamically unstable mass loss.\footnote[3]{We 
should mention here, in a broader context, the importance of an accurate 
consideration of the stellar wind mass loss before RLOF for the predicted period 
distribution of the population of double degenerates.}      

5. The rate of SN~Ia as a function of redshift generally follows the
star formation rate curve, with a small shift of the peak (by about
$\Delta z \sim 0.1$) towards lower redshifts.  The ratio of the rates of
SN~Ia to SN~II, Ib, Ic stays flat for $z \apgt 1$, but increases from $z
\sim 1$ towards lower redshifts, and it is higher at $z \sim 0$, by
about a factor 4, than its value at $z \sim 1$.  A similar trend is expected 
for the nucleosynthetic abundances ratio of, for example, [Fe/O], but great
care should be taken in the interpretation of actual measurements (see
e.g.\ Kulkarni, Fall, \& Truran 1997).

\bigskip

\begin{acknowledgments}

We a grateful to A.~Fedorova for the computation of several representative 
stellar models upon our request.
Helpful discussions with A.~Tutukov, P.~Madau,  and P.~Kahabka and a 
critical reading of the manuscript by E.~P.~J.~van den Heuvel and the
referee, P.~Ruiz-Lapuente, 
are acknowledged. LY acknowledges the support and hospitality of the
ST~ScI and support from RFBR grant 96-02-16351.
ML acknowledges support from NASA Grant NAGW-2678.

\end{acknowledgments}
 
\pagebreak

\pagebreak

\noindent{\bf Figure 1.}~The dependence
of the derivatives of the Roche lobe radius $\xi_L$ on the mass ratio of
the components, for retention efficiencies $\alpha = 1$ and 0 (solid curves).
The dashed lines give the values of the adiabatic response of
stellar radii $\xi_{ad}$, for core mass fractions in condensed polytropes
(Hjellming \& Webbink 1987) $m_c = 0.1962$  and 0.3833.

\noindent{\bf Figure 2.}~The initial masses of accretors and donors in
systems which produce SN~Ia. Plotted is the value of 
$\frac{\partial^2\nu}{\partial M_a\partial M_d}$, where $\nu$ is the
 birthrate of the parent MS system. Upper panel---the HKN model for
\ace=10, lower panel---the HKN model as modified in the present paper
(see text) with a scaling factor for the mass loss rate of $\alpha_{th}$ = 
0.5. The maximum of the gray scale is the same for both panels and 
corresponds to $\frac{\partial^2 \nu}{\partial M_a \partial M_d} = 1.6 
\times 10^{-3}$.

\noindent{\bf Figure 3.}~The dependence of the donor and accretor masses 
on time in systems with $a_0 = 10 \rs$\ (upper curve) and $20 \rs$\ (lower 
curve) in the model with \at = 0.5.

\noindent{\bf Figure 4.}~Upper panel---The contribution of stars of
different ages to the present rate of accumulations of \mc\ (solid
histogram) and of edge-lit detonations (dashed histogram). Lower 
panel---The time evolution of these events under the assumption of constant
star formation rate.

\noindent{\bf Figure 5.}~The dependence of the cosmic SN rate on $z$ (panel 
{\it a}) and on look-back time (panel {\it b}). Long-dashed curve---sum of 
SN~II, Ib, and Ic descending from massive stars; short-dashed line---accumulations 
of 0.15 \ms\ of He on top of a white dwarf at an accretion rate of $\md \aplt 3 
\times 10^{-8}$~\myr in systems with nondegenerate helium donors; solid
line---mergers of double degenerates with a total mass greater than \mc; dotted
line in panel {\it a} gives the logarithm of the ratio of the rate of mergers of 
double degenerates to the rate of explosions of massive stars (a factor $-4.5$ was 
added for scaling).

\pagebreak
\centerline{\bf TABLE 1. Occurrence Rates of SN Ia and }
\centerline{\bf  Related Events Under Different Model Assumptions }    
\small\tabcolsep=7pt
\centerline{
\begin{tabular}{rrllllrr}
\hline
\hline
N & \ace & $\alpha_{th}$ & SN Ia & AIC & ELD & $N_{obj}$ & $R_N$ \\
  &      &               & (\pyr) & (\pyr) & (\pyr) & & (\pyr) \\ 
(1) & (2) &         (3)  & (4)    & (5) & (6) & (7) & (8) \\
\hline
1 & 1 & -- & 0.0 & 0.0 & 0.0 & 30 & 5  \\
2 & 2 & -- & 0.0 & $8 \times 10^{-7}$ & $3 \times 10^{-6}$ & 50 & 9 \\
3 & 5 & -- & $3 \times 10^{-7}$ & $5 \times 10^{-6}$ & $5 \times 10^{-6}$ & 80 &
21 \\
4 & 10 & -- & $2 \times 10^{-6}$ & $2 \times 10^{-5}$ & $6 \times 10^{-7}$ & 110
& 27 \\
5 & 20 & -- & $5 \times 10^{-6}$ & $4 \times 10^{-5}$ & 0.0 & 130 & 32 \\
\hline
6 & 1 & 1 & $7 \times 10^{-5}$ & $5 \times 10^{-5}$ & $3 \times 10^{-3}$ & 6100
& 31 \\
7 & 1 & 0.5 & $2 \times 10^{-4}$ & $4 \times 10^{-5}$ & $3 \times 10^{-3}$ & 8500
& 50 \\
8 & 1 & 0.2 & $1 \times 10^{-4}$ & $2 \times 10^{-5}$ & $4 \times 10^{-3}$ & 9700
& 110 \\
9 & 1 & 0.1 & $1 \times 10^{-5}$ & $7 \times 10^{-5}$ & $4 \times 10^{-3}$ &
9400 & 84 \\
\hline
10 & 1 & -- &   0.0              & $8 \times 10^{-6}$ & $1 \times 10^{-5}$ &
400   & 12 \\                                                                       
11 & 10 & -- &  0.0               & $2 \times 10^{-5}$ &  0.0               &
220   & 25 \\                                                                          
\hline
12 & 1 & 1 & 0.0 & 0.0 & 0.0 & 37 & 3 \\
\hline
13 & 1 & 1 & $9 \times 10^{-5}$ & $7 \times 10^{-5}$ & $5 \times 10^{-3}$ &
11000 & 74 \\                                                                       
14 & 1 & 0.5 & $3 \times 10^{-4}$ & $8 \times 10^{-5}$ & $6 \times 10^{-3}$ &
16000 & 110 \\
15 & 1 & 0.2 & $1 \times 10^{-3}$ & $5 \times 10^{-5}$ & $7 \times 10^{-3}$ &
20000 & 74 \\                                                                       
16 & 1 & 0.1 & $8 \times 10^{-4}$ & $2 \times 10^{-5}$ & $9 \times 10^{-3}$ &
25000 & 380 \\
\hline
\end{tabular}
}%end centerline

\medskip

Notes: N---number of model; \ace---common envelope parameter; 
$\alpha_{th}$---scaling factor for mass transfer rate; SN~Ia---occurrence 
rate of SN~Ia via the accumulation of \mc; AIC---occurrence rate of accretion 
induced collapses; ELD---occurrence rate of edge-lit detonations after the 
accumulation of 0.15~\ms\ of He (see text); $N_{obj}$---Galactic number of  
objects; $R_N$---occurrence rate of hydrogen flashes in the nuclear burning 
controlled evolutionary stage.

\pagebreak
\begin{figure}
\centerline{\psfig{file=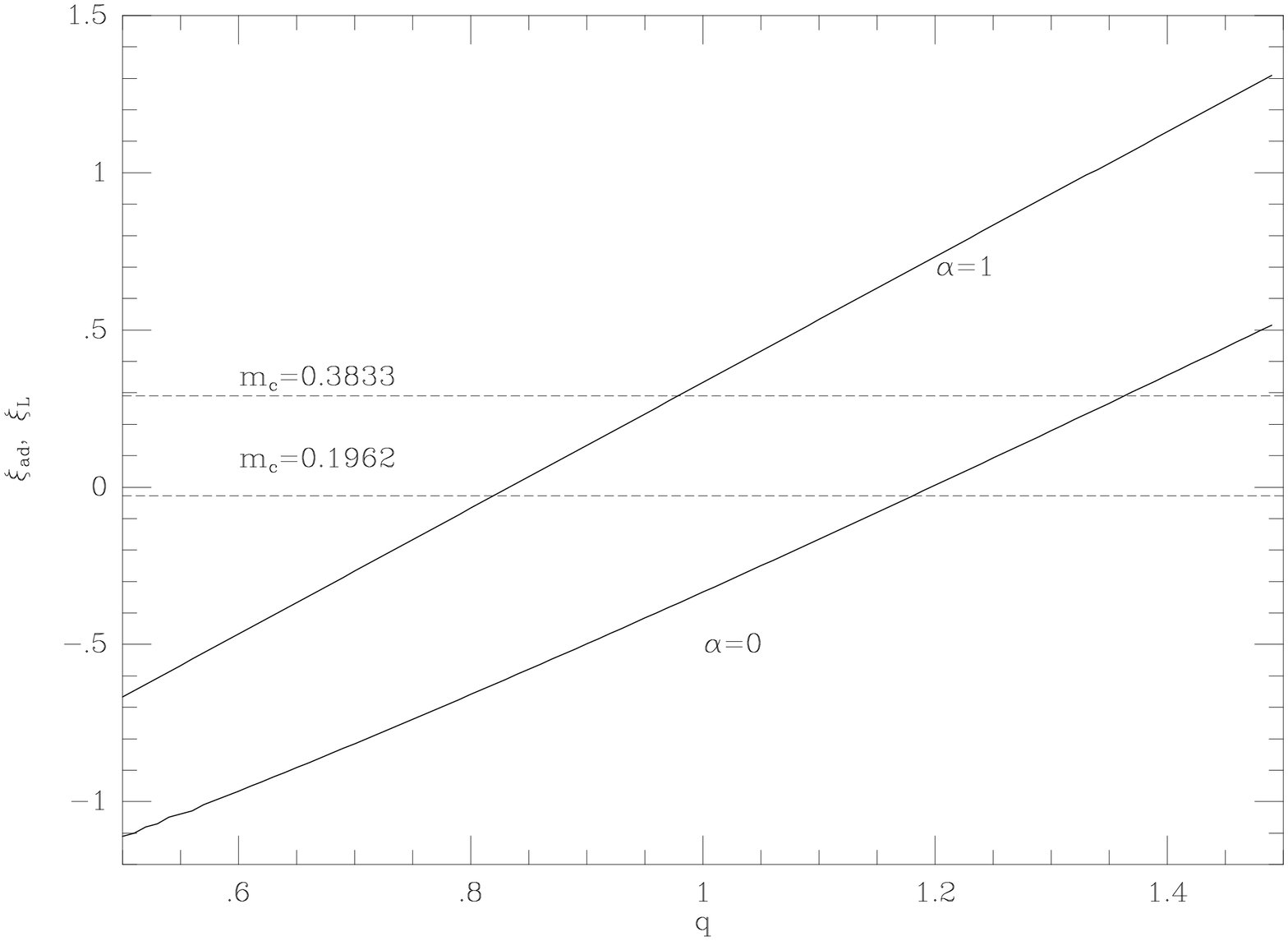,bbllx=50pt,bblly=50pt,bburx=500pt,bbury=700pt,width=16 cm,angle=-90}}
\vspace{2cm}
\caption[]{ }
\end{figure}
\pagebreak
\vspace{-6cm}
\begin{figure}
\centerline{\psfig{file=ylfig2.ps,angle=-90}}
\vspace{0.5cm}
\caption[]{ }
\end{figure}

\begin{figure}
\centerline{\psfig{file=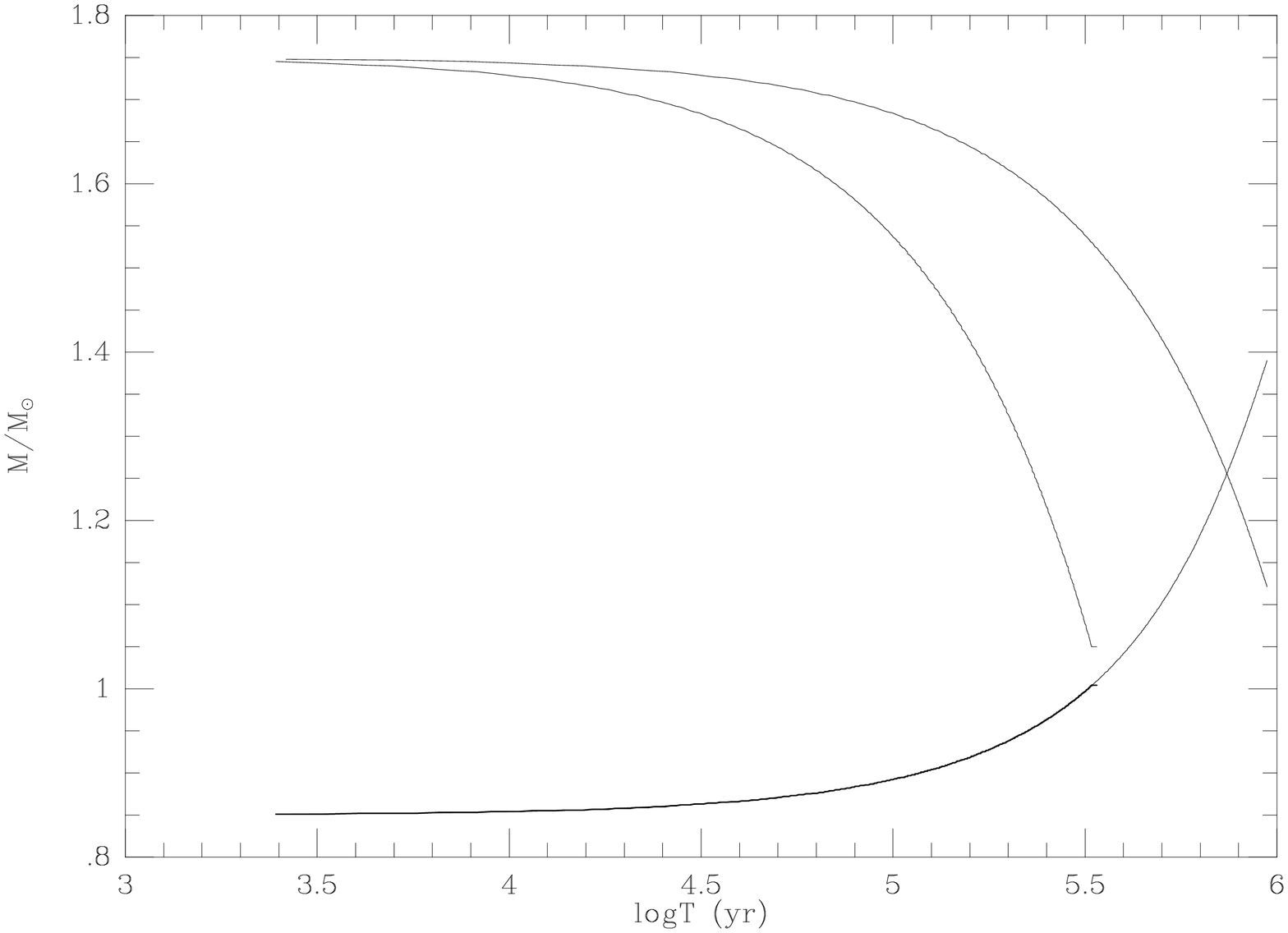,bbllx=50pt,bblly=50pt,bburx=500pt,bbury=700pt,width=16 cm,angle=-90}}
\vspace{2cm}
\caption[]{ }
\end{figure}
\pagebreak
\begin{figure}
\vspace{-3cm}
\centerline{\psfig{file=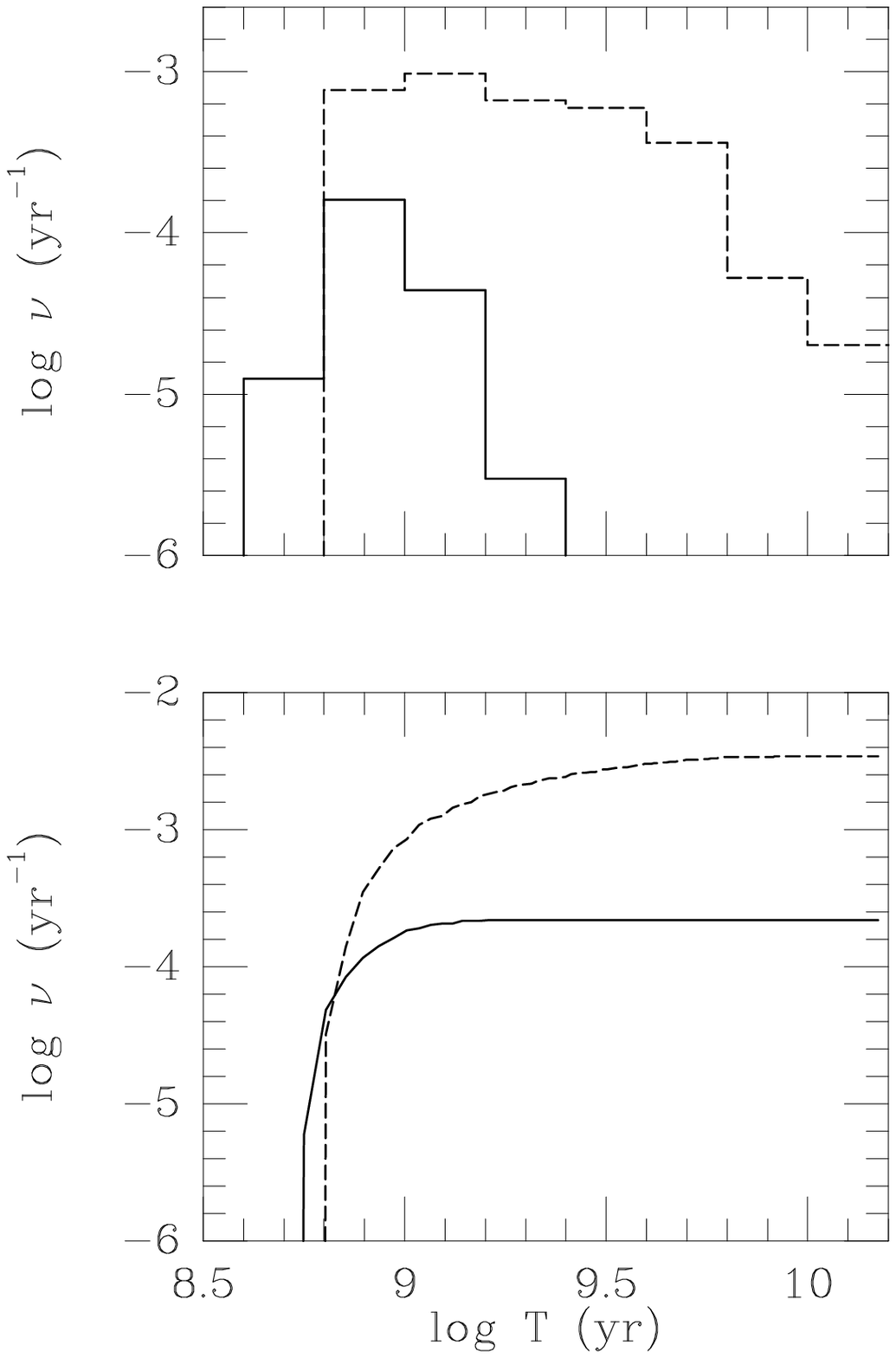,bbllx=50pt,bblly=50pt,bburx=500pt,bbury=700pt,width=16 cm}}
\vspace{-3cm}
\caption[]{ }
\end{figure}
\pagebreak
\begin{figure}
\centerline{\psfig{file=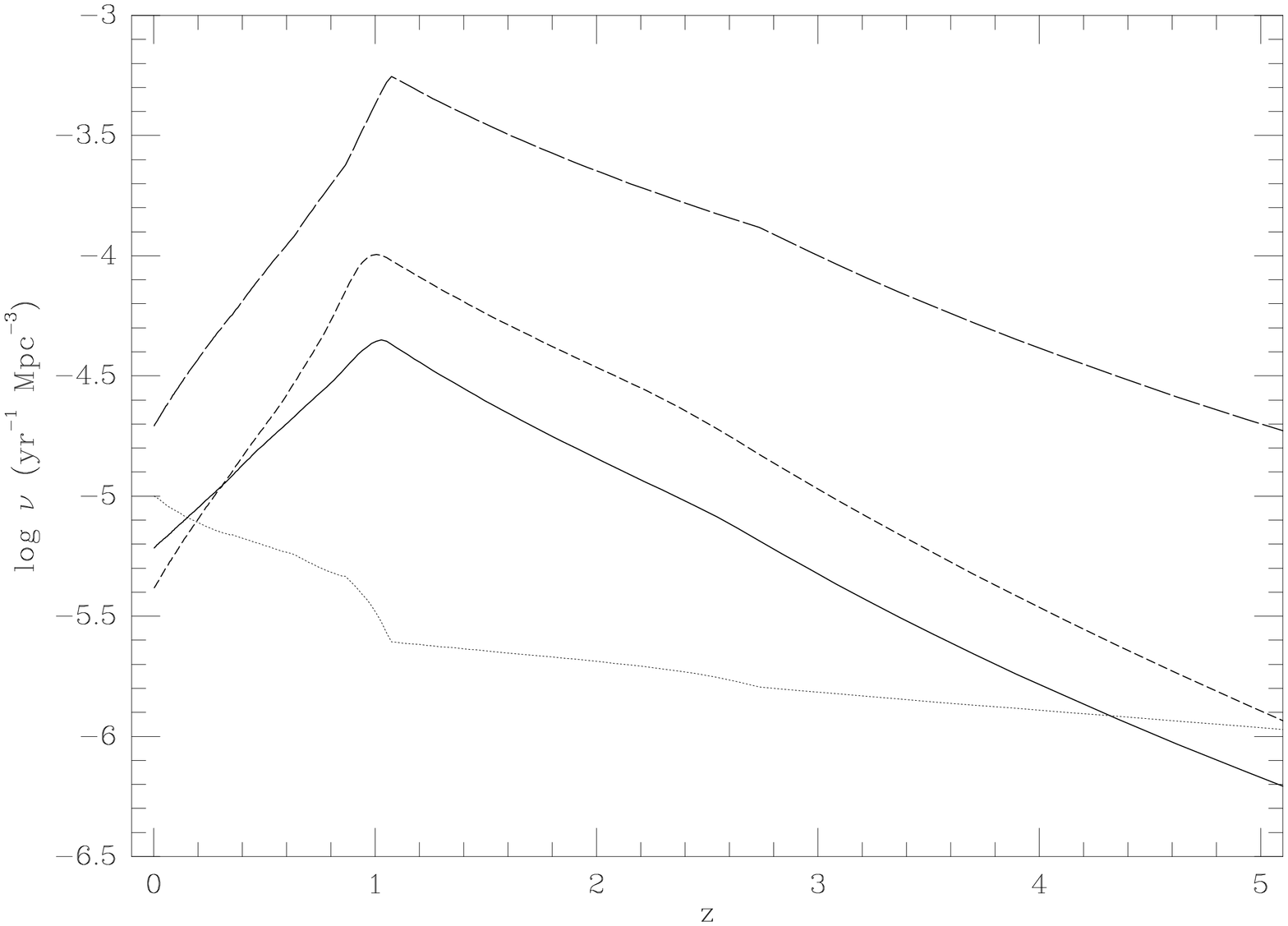,bbllx=50pt,bblly=50pt,bburx=500pt,bbury=700pt,width=16 cm,angle=-90}}
\vspace{2cm}
%\caption[]{ }
Figure 5a:
\end{figure}
\pagebreak
\begin{figure}
\centerline{\psfig{file=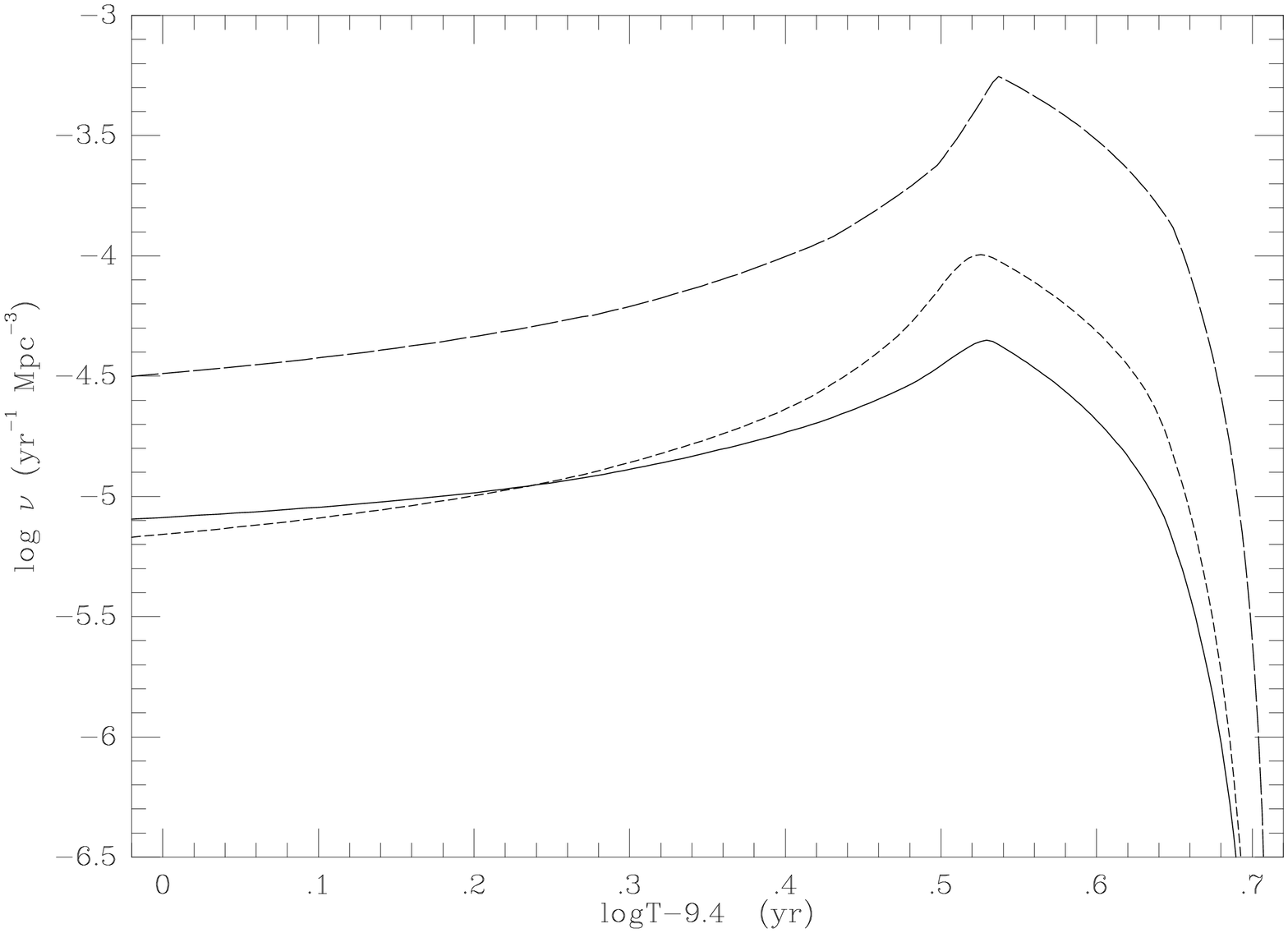,bbllx=50pt,bblly=50pt,bburx=500pt,bbury=700pt,width=16 cm,angle=-90}}
\vspace{2cm}
%\caption[]{ }
Figure 5b:
\end{figure}
\end{document}